\documentclass[sigconf]{acmart}

\usepackage{enumitem}
\usepackage[ruled,linesnumbered,noend]{algorithm2e}
\usepackage{setspace}

\SetCommentSty{mycommfont}

\newcommand{\tabledL}[1]{\textcolor[rgb]{0.215, 0.494, 0.722}{\textbf{#1}}}
\newcommand{\tablet}[1]{\textcolor[rgb]{1.0, 0.498, 0.0}{\textbf{#1}}}

\AtBeginDocument{%
  \providecommand\BibTeX{{%
    \normalfont B\kern-0.5em{\scshape i\kern-0.25em b}\kern-0.8em\TeX}}}
\setcopyright{acmcopyright}
\copyrightyear{2023}
\acmYear{2023}
\acmDOI{XXXXXXX.XXXXXXX}

\acmConference[KDD 2023]{Make sure to enter the correct
  conference title from your rights confirmation email}{Aug 06--10,
  2023}{Long Beach, CA}
\acmPrice{15.00}
\acmISBN{978-1-4503-XXXX-X/18/06}

\begin{document}

\title{\textsc{Beep}: Balancing Effectiveness and Efficiency when Finding Multivariate Patterns in Racket Sports}
\settopmatter{authorsperrow=4}

\author{Jiang Wu}
\email{wujiang5521@zju.edu.cn}
\affiliation{%
  \institution{Zhejiang University}
  \city{Hangzhou}
  \state{Zhejiang}
  \country{China}
}

\author{Dongyu Liu}
\email{dyuliu@ucdavis.edu}
\affiliation{%
  \institution{University of California, Davis}
  \city{Davis}
  \state{California}
  \country{USA}
}

\author{Ziyang Guo}
\email{ziyangguo27@zju.edu.cn}
\affiliation{%
  \institution{Zhejiang University}
  \city{Hangzhou}
  \state{Zhejiang}
  \country{China}
}

\author{Yingcai Wu}
\authornote{Yingcai Wu is the corresponding author.}
\email{ycwu@zju.edu.cn}
\affiliation{%
  \institution{Zhejiang University}
  \city{Hangzhou}
  \state{Zhejiang}
  \country{China}
}

\renewcommand{\shortauthors}{Wu et al.}

\begin{abstract}
  Modeling each hit as a multivariate event in racket sports and conducting sequential analysis aids in assessing player/team performance and identifying successful tactics for coaches and analysts.
  However, the complex correlations among multiple event attributes require pattern mining algorithms to be highly effective and efficient.
  This paper proposes \textsc{Beep} to discover meaningful multivariate patterns in racket sports.
  In particular, \textsc{Beep} introduces a new encoding scheme to discover patterns with correlations among multiple attributes and high-level tolerances of noise.
  Moreover, \textsc{Beep} applies an algorithm based on LSH (Locality-Sensitive Hashing) to accelerate summarizing patterns.
  We conducted a case study on a table tennis dataset and quantitative experiments on multi-scaled synthetic datasets to compare \textsc{Beep} with the SOTA multivariate pattern mining algorithm.
  Results showed that \textsc{Beep} can effectively discover patterns and noises to help analysts gain insights.
  Moreover, \textsc{Beep} was about five times faster than the SOTA algorithm.
\end{abstract}





\maketitle

\begin{spacing}{0.985}
\section{Introduction}

\label{sec:Introduction}

Multivariate event sequence data is widely analyzed in racket sports, such as table tennis \cite{wang2019tac, wang2021tac, Lan2021RallyComparator}, tennis \cite{polk2019courttime,wu2022rasipam}, and badminton \cite{wu2020visual,wu2021tacticflow}.
These works usually model each hit as an event with multiple attributes, each for a hitting feature (e.g., the ball position, the hitting technique, etc.).
Consecutive hits, starting with one player serving the ball and ending with one player winning a point, constitute a multivariate event sequence.
Based on such a data model, pattern mining algorithms can discover frequent patterns (shown as Figure \ref{fig:patterndef}), which can be regarded as players' tactics, to help domain experts in sports obtain insights into players' playing styles and thereby improve their performance.
However, in five years of collaboration with domain experts in racket sports, we found that multivariate event sequences placed high-level requirements on both \textbf{effectiveness} and \textbf{efficiency} of pattern mining algorithms.

\textbf{Effective} pattern mining algorithms can discover patterns that reveal meaningful information about the sequences. To be effective, a pattern mining algorithm for multivariate event sequences must fulfill three conditions:
(1) The correlations among multiple attributes within sequences should be preserved for domain analysis.
For example, in table tennis, a tactical pattern may be that when a player hits the ball to a certain \textit{position} on the table, the opponent always uses a specific \textit{technique} in response.
(2) The algorithm should have a high tolerance for single-value noises (i.e., changes on only one attribute).
For example, in table tennis, when a player applies a tactical pattern, he/she may only change the technique of one hit to a similar technique, retaining the overall playing style.
(3) The number of returned patterns should be manageable.
Given that multivariate patterns are complicated, analyzing them is both time-consuming and mentally overwhelming.

\textbf{Efficient} pattern mining algorithms can mine multivariate patterns within an acceptable response time.
In practice, the time allowed for pattern analysis is limited (e.g., one hour).
Moreover, some parameters should be adjusted based on the analysts' feedback to satisfy the analysts' requirements.
Thus, the algorithm is expected to return the results in several minutes.

To the best of our knowledge, existing multivariate pattern mining algorithms cannot satisfy these two requirements simultaneously.
Some algorithms \cite{morchen2007efficient,chen2010efficient,bertens2014characterising} transform multivariate sequences into univariate ones, such that extracted patterns cannot retain any correlations between attributes.
Algorithms based on SPM (Sequential Pattern Mining) \cite{oates1996searching,tatti2011mining,wu2013mining,fournier2017survey} retain the correlations between attributes but usually return an enormous number of patterns rather than seeking the most meaningful ones, due to the well-known problem of \textit{pattern explosion} \cite{menger2015experimental}.
Algorithms based on MDL (Minimum Description Length) summarize a set of patterns to describe the entire sequence dataset instead of searching for each pattern \cite{bertens2016keeping,kawabata2018streamscope,wu2020visual}, thus avoiding \textit{pattern explosion}.
However, the current MDL-based methods have no tailored algorithmic design to handle single-value noises in sports and are usually time-consuming.

In this paper, we propose \textsc{Beep}, a novel pattern mining algorithm which \textbf{B}alances \textbf{E}ffectiveness and \textbf{E}fficiency when finding \textbf{P}atterns in racket sports.
The contributions are mainly as follows.
\begin{itemize}[leftmargin=10pt]
    \item We introduce a new encoding scheme for the noise values in a pattern, enhancing \textsc{Beep}'s effectiveness (i.e., tolerance of noises).
    \item We propose a tailored acceleration method based on \textit{Locality Sensitive Hashing} so that the patterns with high frequencies can be found in a short time, enhancing \textsc{Beep}'s efficiency.
    \item We conducted an empirical study with analysts in table tennis to demonstrate that \textsc{Beep} can help analysts obtain insights into players' playing styles.
    We further compared \textsc{Beep} with the current SOTA algorithm on multi-scaled synthetic datasets, proving that \textsc{Beep} was about five times faster than the SOTA algorithm.
\end{itemize}

\section{Problem Formulation}

\label{sec:Predefinitions}

\begin{figure}[!tp]
  \centering
  \includegraphics[width=\linewidth]{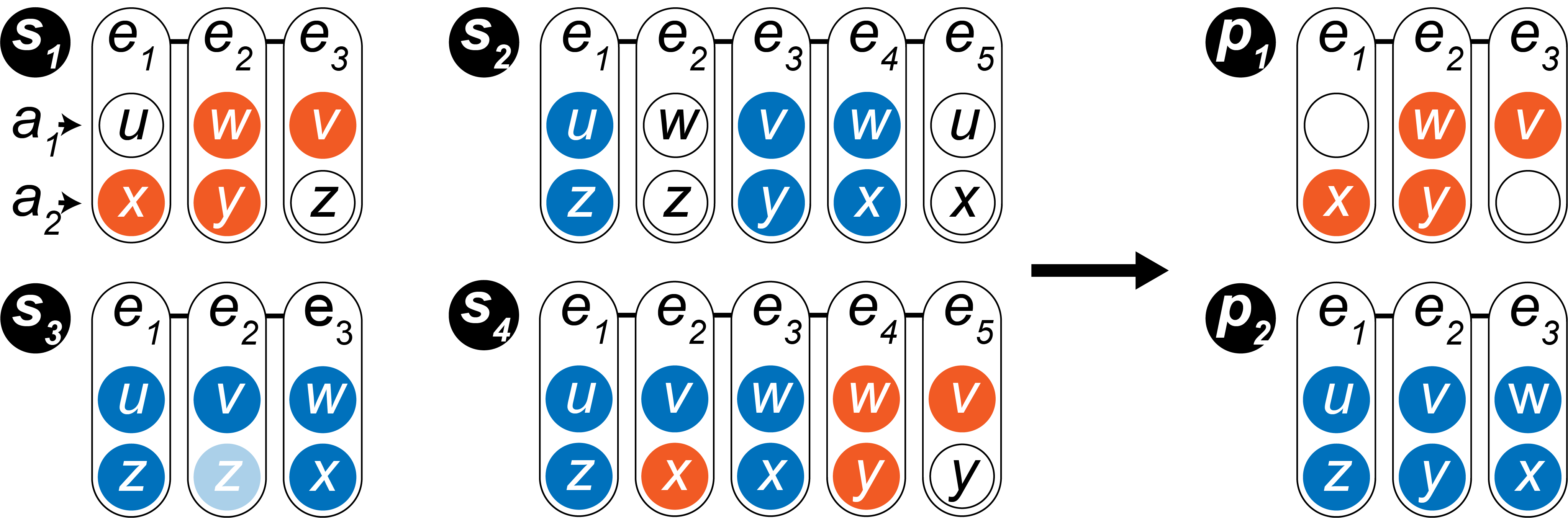}
  \caption{Examples of patterns over two attributes. There exist two patterns ($p_1$ and $p_2$) from four sequences ($s_1$ to $s_4$).}
  \label{fig:patterndef}
\end{figure}

\textsc{Beep} accepts a multivariate event sequence dataset $S$ as input and outputs a pattern set $P$, demonstrated as Figure \ref{fig:patterndef}.

\textbf{Each sequence} $s \in S$ is a vector of events, denoted $s = (e_1, e_2, ...,\\ e_{|s|})$.
All sequences are over the same set of categorical attributes $A=\{a_1, a_2, ..., a_{|A|}\}$.
Thus, each event $e$ can be considered a vector of values, denoted $e=(a_1=v_1, a_2=v_2, ..., a_{|A|}=v_{|A|})$.

\textbf{Each pattern} $p \in P$ is a subsequence of numerous original sequences in the dataset $S$, where $|p|$ indicates the length and $||p||$ indicates the number of non-empty values.
For example, in Figure \ref{fig:patterndef}, four sequences ($s_1$ to $s_4$) can be summarized by pattern $p_1$ ($||p_1||=4$) and pattern $p_2$ ($||p_2||=6$).
The formal definition of ``subsequence'' can be found in Appendix \ref{app:subsequence}.
\textsc{Beep} further supports four features as follows to obtain informative patterns in real-world datasets.

\begin{itemize}[leftmargin=10pt]
  \item \textbf{Empty values} in a pattern can match any values.
  For example, in Figure \ref{fig:patterndef}, $v_2$ of $e_3$ in $p_1$ is empty, which matches $z$ in $s_1$ and $y$ in $s_4$.
  Empty values allow \textsc{Beep} to tolerate the ever-changing values in a pattern, which were less important.

  \item \textbf{Missing values} allow patterns to tolerate some noise values in sequences.
  For example, $s_3$ misses $v_2$ of $e_2$ when it is covered by $p_2$.
  \textsc{Beep} considers the possibility (common in real-world data) that the missing value is actually a noise value.

  \item \textbf{Gap events} can separate two consecutive events in a pattern into non-consecutive events in sequences.
  For example, sequence $s_2$ has pattern $p_2$, but the gap event $e_2$ of $s_2$ is not captured by $p_2$.
  \textsc{Beep} allows gaps to ignore events that may be noise.

  \item \textbf{Interleaving patterns} overlap over a period of time\cite{tatti2012long}, e.g., pattern $p_1$ and $p_2$ are interleaving in $s_4$.
  \textsc{Beep} allows interleaving patterns to discover simultaneous patterns.
\end{itemize}

In practice, domain experts usually expect patterns to be informative (i.e., fewer empty values), authentic (i.e., fewer missing values), and compact (i.e., fewer gaps).
Thus, for each pattern $p$, we limit the maximum number of empty values ($|p|$), missing values ($||p|| \div 10$, and at most 1 for each event), and gap events ($|p|-1$).

\section{Beep}

In this section, we introduce our algorithm, \textsc{Beep}, to mine multivariate patterns in racket sports.
Based on the MDL principle \cite{grunwald2007minimum}, \textsc{Beep} describes the dataset into two parts -- i.e., the pattern set and the description of each sequence based on the pattern set -- and expects the minimum description of the two parts (Equation \ref{eq:mdl}).
\begin{equation}
    \begin{aligned}
    \mathop{minimize}\limits_{P} \ L(S~|~P),~where~L(S~|~P) = L(P) + \sum_{s\in S}L(s~|~P),\\
    \end{aligned}
    \label{eq:mdl}
\end{equation}
where $L(S~|~P)$, $L(P)$, and $L(s~|~P)$ is the description length of dataset $S$ based on $P$, pattern set $P$, and sequence $s$ based on $P$, respectively (detailed calculation can be found in Section \ref{sec:coverAlgorithm}).

Inspired by Ditto \cite{bertens2016keeping}, \textsc{Beep} optimizes $L(S~|~P)$ iteratively and heuristically.
Initially, $P$ includes all singleton patterns (i.e., an event with only one non-empty value).
At each iteration, \textsc{Beep} generates a set of candidate patterns $C$ and filters each usefull candidate $cp \in C$ (i.e., $L(S~|~P + cp) < L(S~|~P)$).
After adding these useful candidates into $P$, \textsc{Beep} further filters out each redundant pattern $p \in P$ (i.e., $L(S~|~P - p) < L(S~|~P)$).
\textsc{Beep} will not stop until $P$ keeps the same at a certain iteraction, which indicates that $P$ approximates the optimal pattern set with the minimum description length $L(S~|~P)$.
The key contribution of \textsc{Beep} lies in three aspects as follows.

\subsection{Candidates Generation}

\label{sec:candidates}

\begin{figure}
    \centering
    \includegraphics[width=\linewidth]{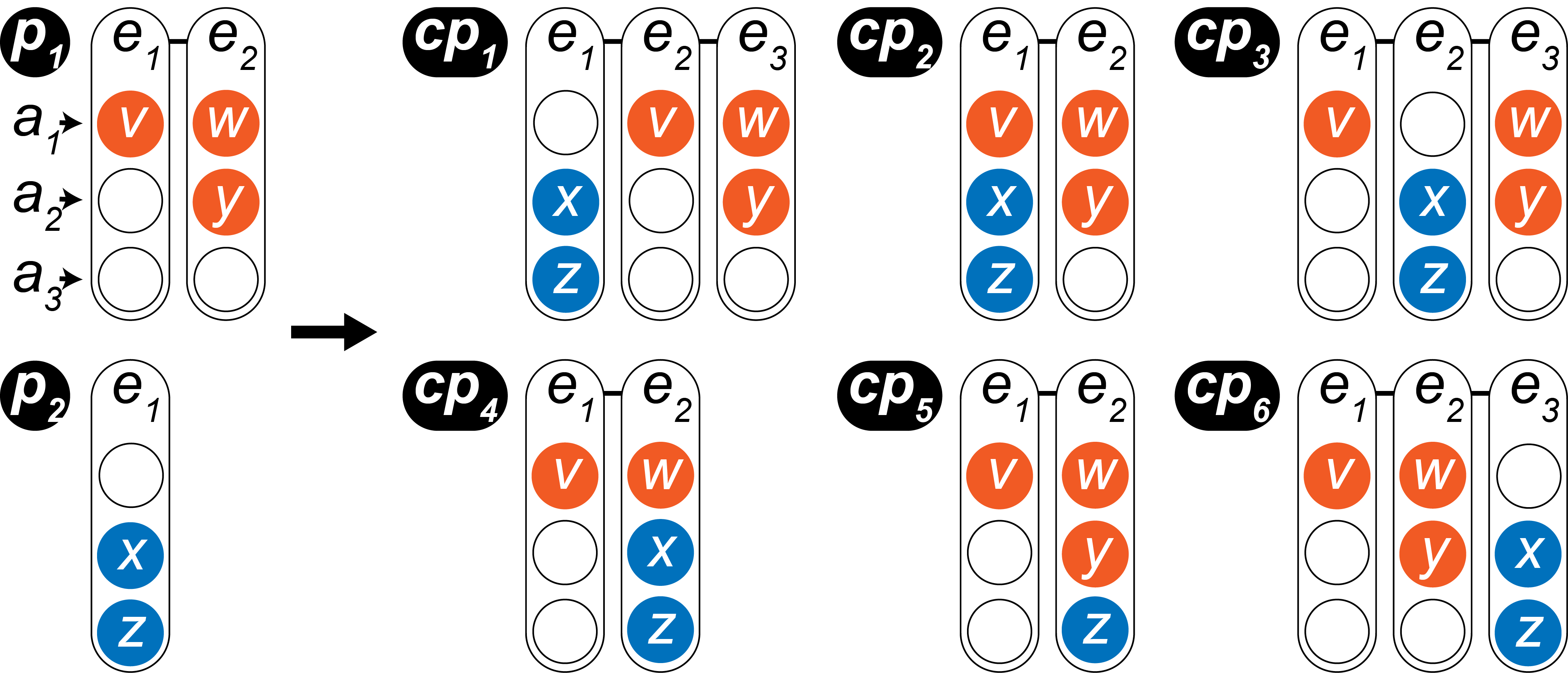}
    \caption{Examples for constructing candidate patterns. Given two patterns $p_1$ and $p_2$, we construct six candidate patterns $cp_1$ to $cp_6$ at different alignments. Candidates $cp_4$ and $cp_5$ demonstrate how we fix a conflict by missing values.}
    \label{fig:candidate}
\end{figure}

\textsc{Beep} constructs complex patterns with simple ones, i.e., combining each pair of patterns in $P$ at different alignments.
For example, in Fig. \ref{fig:candidate}, we list all the candidate patterns ($cp_1$ to $cp_6$) generated based on multivariate pattern $p_1$ and $p_2$ over three attributes.
However, when aligning $e_2$ of $p_1$ and $e_1$ of $p_2$, a conflict exists on value $x$ and $y$.
We resolve this conflict by regarding it as a missing value of one of the original patterns.
For example, a sequence perfectly matched by $p_1$ may also be covered by $cp_4$ with one missing value.

\subsection{Description Length Calculation}

\begin{figure}[!tp]
    \centering
    \includegraphics[width=\linewidth]{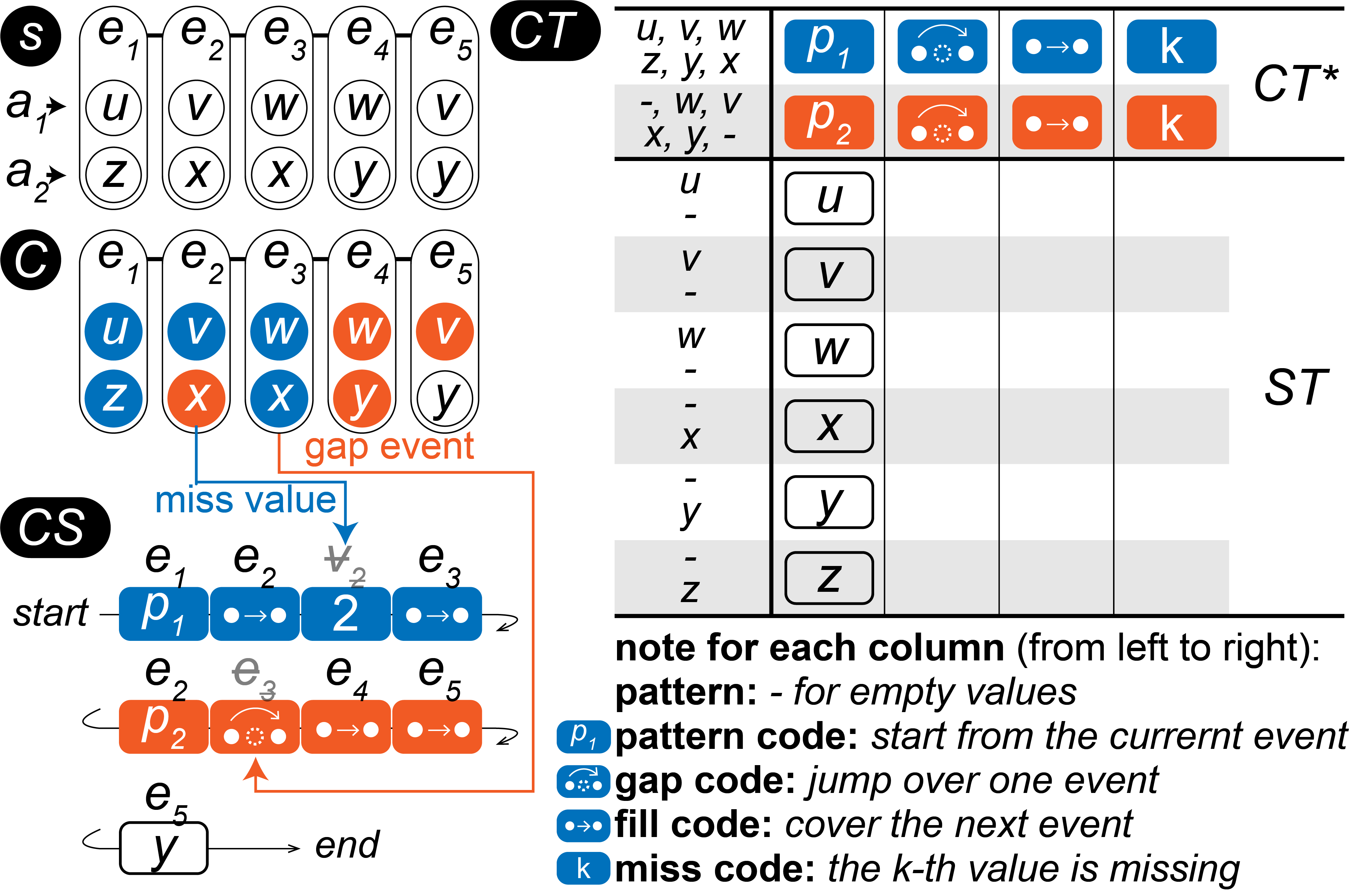}
    \caption{Examples for explaining the process of encoding. Given sequence $s$ and code table $CT$, we first obtain cover $C$ by describing $s$ with the patterns in $CT$. We further obtain the code stream $CS$, where $s$ is encoded by the codes in $CT$.}
    \label{fig:ct}
\end{figure}

\label{sec:coverAlgorithm}

The process of calculating description length is shown as Figure \ref{fig:ct}, where we convert pattern set $P$ into a code table ($CT$) and use $CT$ to encode each sequence $s\in S$ into a code stream $CS$.
We define $L(P)$ and $L(s | P)$ as the digital length of $CT$ and $CS$, respectively.
Detailed mathematical calculations can be found in Appendix \ref{app:mdl}.

\subsubsection{Code table}
The code table records a pattern code for all patterns in $P$, including the generated patterns ($CT^*$) and singleton patterns ($ST$).
For the generate patterns, the code table further records its gap code (for encoding gap events), fill code (for filling in the next event in $p$), and miss code (for encoding missing values).
We employ the Huffman coding algorithm \cite{moffat2019huffman} to represent each code as a distinct binary number according to the frequency it is used to encode the sequences in $S$, in order to ensure that the total description length is minimal.

\subsubsection{Sequence encoding}
Given a code table $CT$ and a sequence $s$, we first cover $s$ using the patterns in $CT$ ($C$ in Figure \ref{fig:ct}, pseudocode is in Appendix \ref{app:search}).
Then, \textsc{Beep} scans each value in $s$ in order and pushes the corresponding code to $CS$.
For example, sequence $s$ in Fig. \ref{fig:ct} is covered by pattern $p_1$, pattern $p_2$ and singleton value $y$.
Pattern $p_1$ starts from $e_1$ (the pattern code) and covers $e_2$ and $e_3$ (two fill codes), where $v_2$ of $e_2$ is missing (the miss code).
Pattern $p_2$ starts from $e_2$ (the pattern code), gaps at $e_3$ (the gap code), and covers $e_4$ and $e_5$ (two fill codes).
Singleton $y$ occurs at the last.

\subsection{Algorithm Acceleration}

\label{sec:lsh}

Description length calculation is the most time-consuming step, as it runs for each of thousands of candidates and considers thousands of original sequences (see Appendix \ref{app:search} for time complexity analysis).
However, a candidate $c$ is generated by combining each two patterns.
Only when two patterns often occur simultaneously or consecutively, the combined candidate is more likely to be a frequent pattern and benefits description length reduction.

Thus, we accelerate \textsc{Beep} by filtering out those pairs of patterns with low co-occurrence in advance.
For each pattern $p$, we record the positions where it occurs when covering the dataset.
A position is an index of a segment of original sequences, where a segment contains at most $l$ events (default $l$ is 20), and a long sequence will be cut into several segments.
Then, we apply weighted \textit{Locality Sensitive Hashing} (LSH) \cite{ioffe2010improved} to examine whether the two patterns occur in similar sets of positions.
If two sets of numbers have a weighted Jaccard similarity larger than a threshold $th$, the weighted LSH algorithm will regard them as the same.

\section{Experiments}

\label{sec:Experiments}

\begin{figure}
    \centering
    \includegraphics[width=\linewidth]{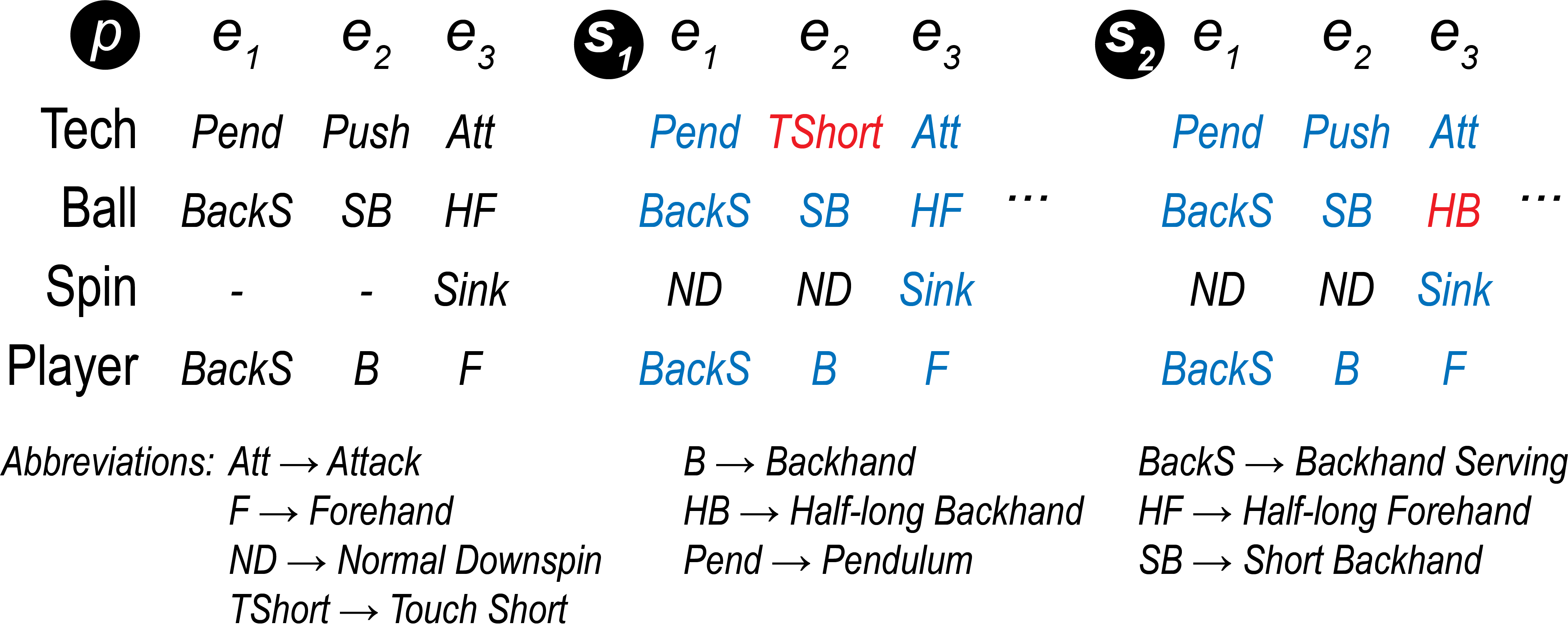}
    \caption{Patterns found in table tennis data. Two sequences $s_1$ and $s_2$ contain pattern $p$ (highlighted in blue) and have missing values (highlighted in red).}
    \Description{Summarizing two patterns from four sequences over two attributes.}
    \label{fig:tt}
\end{figure}

\begin{table}
    \caption{The results of the case study on the table tennis data. We compared two algorithms, namely \textsc{Ditto} and \textsc{Beep}, on the number of patterns ($|P|$), the average frequency of each pattern ($avg.$ $freq.$), and the number of missing values in 716 sequences ($miss$, only for \textsc{Beep}), and the runtime ($t$).}
    \label{tab:empirical}
    \begin{tabular}{rrrrr}
        \toprule

        \textit{Algorithm} & $|P|$ & $avg.$ $freq.$ & $miss$ & $t$(s) \\

        \midrule

        \textsc{Ditto} & 161 & 4.36 & -- & 1747 \\

        \textsc{Beep} & 62 & 12.94 & 280 & 425 \\

        \bottomrule
    \end{tabular}
\end{table}

\begin{table*}
    \caption{The results of the quantitative experiments. Given 5 synthetic datasets, one for each row, we compared five algorithms, namely, \textsc{Ditto}, \textsc{Beep}, \textsc{Beep}-miss, \textsc{Beep}-LSH, and \textsc{Beep}-none. For each dataset, we reported the number of sequences ($|S|$), the length of each sequence ($|s|$), the number of attributes ($|A|$), and the number of optional values of each attribute ($|V_k|$). For each algorithm, we reported the number of patterns ($|P|$), the description length reduction compared to the singleton-only model ($\Delta L\%$), the number of missing values for \textsc{Beep}-miss and \textsc{Beep} ($miss$), and the runtime in seconds ($t$). (\tablet{orange}: the shortest time; \tabledL{blue}: the largest description length reduction.)}
    \label{tab:quantitative}
    \begin{tabular}{rrrr|rrr|rrrr|rrrr|rrr|rrr}
      \toprule

      \multicolumn{4}{l|}{Synthetic Datasets}&\multicolumn{3}{l|}{\textsc{Ditto}}&\multicolumn{4}{l|}{\textsc{Beep}}&\multicolumn{4}{l|}{\textsc{Beep}-miss}&\multicolumn{3}{l|}{\textsc{Beep}-LSH}&\multicolumn{3}{l}{\textsc{Beep}-none} \\

      \cmidrule(lr){1-4}\cmidrule(lr){5-7}\cmidrule(lr){8-11}\cmidrule(lr){12-15}\cmidrule(lr){16-18}\cmidrule(lr){19-21}

      $|S|$&  $|s_i|$&      $|A|$&  $|V_k|$&
      $|P|$&  $\Delta L\%$& $t$(s)&
      $|P|$&  $\Delta L\%$& $miss$& $t$(s)&
      $|P|$&  $\Delta L\%$& $miss$& $t$(s)&
      $|P|$&  $\Delta L\%$& $t$(s)&
      $|P|$&  $\Delta L\%$& $t$(s) \\

      \midrule

      50& 20& 5& 100&
      19&   24.5&     399&
      9&   \tabledL{27.2}&     33& 79&
      9&   \tabledL{27.2}&     33&455&
      16&   24.4&     \tablet{43}&
      19&   24.5&     380
      \\

      \textbf{70}& 20& 5& 100&
      12&   19.0&     987&
      14&   \tabledL{21.3}&     11& 215&
      14&   \tabledL{21.3}&     16&1238&
      12&   18.9&     \tablet{29}&
      12&   19.0&     994
      \\

      50& \textbf{30}& 5& 100&
      20&   25.8&     772&
      13&   26.6&     28& 402&
      14&   \tabledL{26.7}&     29& 1276&
      20&   25.9&     \tablet{120}&
      20&   25.8&     753
      \\

      50& 20& \textbf{7}& 100&
      15&   27.7&     1040&
      9&   29.1&     30& 251&
      9&   \tabledL{29.2}&     31& 1141&
      14&   27.7&     \tablet{43}&
      15&   27.7&     1028
      \\

      50& 20& 5& \textbf{200}&
      17&   24.1&     175&
      9&   \tabledL{25.3}&     39& \tablet{18}&
      10&   25.2&     39& 149&
      16&   24.1&     19&
      17&   24.1&     168
      \\

      \bottomrule
    \end{tabular}
  \end{table*}

We implemented \textsc{Beep} in C++\footnote{https://github.com/BEEP-algorithm/BEEP-algorithm} and conducted all experiments on a computer with a 2 GHz CPU and 16 GB of memory.
We mainly conducted two studies to compare \textsc{Beep} with \textsc{Ditto}, one of the SOTA multivariate pattern mining algorithms.

\subsection{Case Study}

To examine that \textsc{Beep} could discover meaningful patterns (for effectiveness), we conducted a case study on a real-world table tennis dataset, with two analysts, each with over three years of experience analyzing table tennis data.
The dataset collected 716 sequences (average length was 5.91) from 10 matches played in 2019 by \textit{Ito Mima}, one of the top players in the world, against Chinese players.
Shown as Figure \ref{fig:tt}, the dataset considered four attributes; namely, the technique the player used to hit the ball (\textit{Tech}), the area where the ball impacted the table (\textit{Ball}), the spin of the ball (\textit{Spin}), and the position of the player (\textit{Player}) (more details can be found in Appendix \ref{app:values}).
We ran both \textsc{Ditto} and \textsc{Beep} on this dataset and summarized the results as follows.

\textbf{\textsc{Beep} can summarize a smaller set of informative patterns in shorter time than \textsc{Ditto}.}
Shown as Table \ref{tab:empirical}, \textsc{Beep} summarized 62 patterns in 425 seconds, while \textsc{Ditto} summarized 161 ones in 1747 seconds.
Meanwhile, the patterns summarized by \textsc{Beep} can be found in more sequences ($avg.$ $freq.$ is 12.94), with a reasonable number of missing values (guaranteed the authentication of patterns).
As a result, \textsc{Beep} reduces the analysis burden of analysts.

\textbf{\textsc{Beep} can find multivariate patterns with tolerances of single-value noises.}
Shown as Figure \ref{fig:tt}, \textsc{Beep} found a multivariate pattern $p$ in two sequences $s_1$ and $s_2$.
Sequence $s_1$ has a missing value at the second hit, where \textit{Ito}'s opponent used the technique \textit{Touch Short}, a control technique similar to \textit{Push}.
The analysts believed that $p$ summarized $s_1$ well, despite the missing value.
Sequence $s_2$ has a missing value at the third hit, where \textit{Ito} received the ball at \textit{Half-long Backhand}, different from \textit{Half-long Forehand} in $p$.
The analysts checked the video and found this value in $s_2$ to be recorded incorrectly.
If we had instead used the \textsc{Ditto} algorithm, these two sequences would have been encoded by other short patterns and thus unable to reveal these insights.

\subsection{Quantitative Experiments}

\label{sec:quantitative}

To quantitatively evaluate the performance of \textsc{Beep}, we compared \textsc{Beep} with \textsc{Ditto} on multi-scaled synthetic data, from the aspects of the number of summarized patterns, the description length reduction (both for evaluating effectiveness), and the runtime (for evaluating efficiency), shown as Table \ref{tab:quantitative}.

Given that \textsc{Beep} mainly introduces two improvements -- namely, the miss codes and the LSH-based acceleration -- we performed an ablation study to evaluate the effects of each improvement.
Specifically, we considered five algorithms: (1) \textsc{Ditto}, (2) the complete \textsc{Beep} algorithm, (3) \textsc{Beep}-miss (only with miss codes), (4) \textsc{Beep}-LSH (only with LSH-based acceleration), and (5) \textsc{Beep}-none (without neither improvements).

We construct 5 datasets, varying in terms of the number of sequences ($|S|$), the length of each sequence ($|s_i|$), the number of attributes ($|A|$), and the number of optional values of each attribute ($|V_k|$).
We further randomly generated \textbf{5 patterns} and planted them to the datasets, each with 5 non-empty values and a random length and covering 10\% of the events in the dataset.
For each pattern, we randomly chose two occurrences and set one value as a missing value for each occurrence, resulting in \textbf{10 total missing values}.
According to our manual check, all five algorithms found all five planted patterns.
Meanwhile, \textsc{Beep}-miss and \textsc{Beep} detected all the planted missing values.

\textbf{\textsc{Beep} and \textsc{Beep}-miss mined a smaller set of patterns with the highest description length reduction.}
Comparing \textsc{Beep}-LSH with \textsc{Beep} and comparing \textsc{Beep}-none with \textsc{Beep}-miss, we find that miss codes can reduce the number of patterns and compress more information.
We believe that miss codes can filter out similar patterns and preserve only one, leaving a smaller set of meaningful patterns.
Miss codes also allow for encoding sequences with long patterns with missing values, rather than multiple short patterns, resulting in a short description length.

\textbf{\textsc{Beep} and \textsc{Beep}-LSH had the lowest runtime.}
Comparing \textsc{Beep}-miss with \textsc{Beep} and comparing \textsc{Beep}-none with \textsc{Beep}-LSH, we found that the LSH-based acceleration substantially reduced runtime but led to more patterns and larger description lengths.

\textbf{In summary, \textsc{Beep} struck a good balance between effectiveness and efficiency and outperformed \textsc{Ditto}.}
\textsc{Beep} balanced the strengths of \textsc{Beep}-miss and \textsc{Beep}-LSH, with a nearly best performance on both information compression and efficiency.
Comparing with \textsc{Ditto}, \textsc{Beep} summarized a more refined set of patterns (smaller $|P|$ in most cases) with less runtime (smaller $t$) and a smaller description length (higher $\Delta L\%$).

\section{Conclusion}

\label{sec:Conclusion}

In this paper, we propose a novel encoding scheme to describe multivariate patterns in event sequence data.
Based on this encoding scheme, we can discover informative patterns that preserve correlations among multiple attributes and highlight any single value noises in the dataset.
To efficiently surface these multivariate patterns from the original sequences, we propose \textsc{Beep}, an MDL-based heuristic algorithm with a tailored acceleration algorithm.
Through a case study on a real-world dataset and quantitative experiments on multi-scaled synthetic data, we prove that our algorithm strikes a good balance between effectiveness and efficiency.

In the future, we plan to integrate more domain knowledge into \textsc{Beep} (e.g., encoding domain knowledge as constraints on the patterns \cite{wu2022rasipam}) to further enhance its effectiveness.
Some subsequences may be frequent (thus summarized by the mining algorithm) but cannot reveal players' playing styles (thus are not worth analyzing).
We also plan to use GPU to accelerate \textsc{Beep} to enhance its efficiency.
Note that covering each sequence with the code table can be parallel.

\begin{acks}
The work was supported by NSFC (U22A2032), and the Collaborative Innovation Center of Artificial Intelligence by MOE and Zhejiang Provincial Government (ZJU).
\end{acks}
\end{spacing}

\bibliographystyle{ACM-Reference-Format}
\bibliography{meta/citations}

\appendix

\section{Covering Algorithm}

\label{app:search}

Algorithm \ref{alg:cover} shows the process of obtaining the optimal cover $C$ given a sequence $s$ and patterns in $CT$.
The core idea is to traverse each pattern $p$ in $CT^*$ (line 2) and try to cover $s$ by $p$ (line 3\textasciitilde7) until all the values in $s$ are marked as covered by a pattern (line 8\textasciitilde9).
If there exist values not marked (line 10\textasciitilde12), they will be marked as covered by the corresponding singleton patterns in $ST$ (line 13\textasciitilde15).
To ensure that the final cover $C$ is optimal, we follow the \textsc{Krimp} algorithm \cite{vreeken2011krimp} and employ a greedy algorithm.
We traverse the patterns in $CT^*$ in a fixed order (Cover Order): $\downarrow ||~p~||$, $\downarrow support(p~|~S)$, and $\uparrow$ lexicographically.
This order ensures that the patterns that have more values and higher frequency, which are more meaningful, will be used first.

\begin{algorithm}[!htb]
    \caption{Covering Algorithm}
    \label{alg:cover}
    \KwIn{A sequence $s$, a code table $CT$}
    \KwOut{An optimal cover $C$}
    \tcc{the explanation can be found in Sec. \ref{sec:coverAlgorithm}}
    $C \leftarrow \emptyset$, $marks \leftarrow \emptyset$, $misses \leftarrow \emptyset$\;
    \For{{\bf each} $p\in CT^*$ in Cover Order} {
        \For{{\bf each} $(marks\_p, misses\_p) \in search(p, s)$}{
            \If{$marks \cap marks\_p = \emptyset$}{
                $C \leftarrow C \cup (marks\_p, misses\_p)$\;
                $marks \leftarrow marks \cup marks\_p$\;
                $misses \leftarrow misses \cup misses\_p$\;
            }
        }
        \If{$|marks| = |s| \times |A|$}{
            {\bf break}\;
        }
    }
    \For{{\bf each} event $e \in s$} {
        \For{{\bf each} value $v \in e$} {
            \If{$v$ is not marked by $marks$} {
                $mark$ = \{cover $v$ by singleton pattern $v$ in $ST$\}\;
                $C \leftarrow C \cup (mark, \emptyset)$\;
                $marks \leftarrow marks \cup mark$\;
            }
        }
    }
    \Return{$C$}
\end{algorithm}

This is a time-consuming algorithm, leading to a high cost for calculating the description length.
The first loop (line 2-9) searches each pattern $p$ in $s$, which is similar to a string matching problem.
We employ the well-known KMP algorithm for searching each pattern, whose time complexity is $O(||s|| + ||p||)$, where $||s||$ and $||p||$ represent the number of values in $s$ and $p$, respectively.
Thus, the total time complexity of the first loop is $O(|P| \times ||s|| + ||P||)$, where $|P|$ and $||P||$ indicate the total number of patterns and values in $P$, respectively.
The second loop (line 10-15) traverses each value in sequence $s$, with a time complexity of $O(||s||)$.
Thus, the time complexity of the covering algorithm is $O(|P| \times ||s|| + ||P|| + ||s||)$.
As we need to cover each sequence $s$ with $CT$ to calculate the description length, the time complexity of the description length calculation is $O(|P| \times ||S|| + ||P|| \times |S| + ||S||)$, where $|S|$ and $||S||$ indicate the total number of sequences and values in $S$, respectively.

\section{Table tennis dataset}
\label{app:values}

There exist four attributes in our table tennis dataset.

\textbf{Tech}: The technique used to hit.
There exist 16 optional values:
\textit{Pendulum serving},
\textit{Reverse serving},
\textit{Tomahawk serving},
\textit{Topspin},
\textit{Smash},
\textit{Attack},
\textit{Flick},
\textit{Twist},
\textit{Push},
\textit{Slide},
\textit{Touch short},
\textit{Block},
\textit{Lob},
\textit{Chopping},
\textit{Pimpled techniques},
\textit{Other techniques},

\textbf{Ball}: The area where the ball hit on the table.
There exist 12 optional values:
\textit{Backhand serving},
\textit{Forehand serving},
\textit{Short forehand},
\textit{Half-long forehand},
\textit{Long forehand},
\textit{Short backhand},
\textit{Half-long backhand},
\textit{Long backhand},
\textit{Short middle},
\textit{Half-long middle},
\textit{Long middle},
\textit{Net or edge},

\textbf{Spin}: The type of ball's spinning.
There exist 7 optional values:
\textit{Strong topspin},
\textit{Normal topspin},
\textit{Strong downspin},
\textit{Normal downspin},
\textit{No spin},
\textit{Sink},
\textit{Without touching},

\textbf{Player}: The position of the player.
There exist 6 optional values:
\textit{Backhand serving},
\textit{Forehand serving},
\textit{Pivot},
\textit{Backhand},
\textit{Forehand},
\textit{Back turn},

\section{Definition of Subsequence}

\label{app:subsequence}

This section introduces the formal definition of ``subsequence'' in Section \ref{sec:Predefinitions} (each pattern $p$ is a subsequence of numerous sequences in $S$).
We introduces the definition in two steps.

First, we define $e^a\preceq e^b$ to indicate that event $e^a$ preserves some attributes of event $e^b$ and drops others, i.e., $e^a$ is part of $e^b$ (e.g., in Figure \ref{fig:patterndef}, $e_1$ of $p_1$, which drops the first attribute, is part of $e_1$ of $s_1$).
Formally, assume that $I$ is the set that includes the indexes of all the attributes to be preserved (i.e., some integers between 1 and $|A|$). For each $1\leq k\leq |A|$, if $k \in I$, $e^a$ has the same $k$-th value as $e^b$.
Otherwise, the $k$-th value of $e^a$ is empty.

Second, we define $s^a \subseteq s^b$ to indicate that sequence $s^a=(e^a_1, ..., e^a_n)$ is a subsequence of sequence $s_b=(e^b_1, ..., e^b_m)$, if there exist integers $1\leq i_1\le i_2\le ...\le i_n\leq m$ such that $e^a_j\preceq e^b_{i_j}$ for each $1\leq j\leq n$.
\section{Description Length Calculation}

\label{app:mdl}

This section provides the complete mathmatical calculation of the description length.

As the basis, we consider the length of each code, which can be computed by Shannon entropy \cite{shannon1948mathematical}.
For pattern $p$, the length of the pattern code $code_p(p)$ is the negative log-likelihood
\begin{displaymath}
    L\left(code_p\left(p\right)\right)=-lg2\left(\frac{usage\left(p\right)}{\sum_{p_i \in CT}usage\left(p_i\right)}\right),
\end{displaymath}
where $lg2(k)$ means the logarithm of $k$ to the base $2$, and $usage(p)$ is the number of pattern codes of pattern $p$ used to encode $S$.
Similarly, we can give the calculation formulas for gap code $code_g(p)$, fill code $code_f(p)$, and miss code $code_m(p)$ as follows.
\begin{displaymath}
    \begin{aligned}
        L\left(code_g\left(p\right)\right)=&-lg2\left(\frac{gaps\left(p\right)}{gaps\left(p\right)+fills\left(p\right)+misses\left(p\right)}\right), \\
        L\left(code_f\left(p\right)\right)=&-lg2\left(\frac{fills\left(p\right)}{gaps\left(p\right)+fills\left(p\right)+misses\left(p\right)}\right), \\
        L\left(code_m\left(p\right)\right)=&-lg2\left(\frac{misses\left(p\right)}{gaps\left(p\right)+fills\left(p\right)+misses\left(p\right)}\right) \\ &+ L_N\left(~|~A~|~\right),
    \end{aligned}
\end{displaymath}
where $gaps(p)$, $fills(p)$, and $misses(p)$ are the number of gap codes, fill codes, and miss codes, respectively, of pattern $p$.
Miss code $code_m(p)$ contains additional bits for the index of missing attributes, which is denoted by $L_N(|A|)$.
Function $L_N(k)$ represents the number of bits required to encode integer $k$, where the MDL optimal Universal code for integers is considered \cite{grunwald2007minimum}.

\textbf{The encoded length of the code table.} To obtain a minimum description length, we treat patterns in $ST$ and patterns in $CT^*$ differently, where $L(P)=L(ST)+L(CT^*)$.

For $ST$, we consider each attribute separately and encode the number of optional values and their supports:
\begin{displaymath}
    L\left(ST\right)=\sum_{1\leq k \leq ~|~A~|~}\left(L_N\left(~|~V_k~|~\right)
    + log\left(
        \begin{array}{c}
                ~|~S^k~|~ \\
                ~|~V_k~|~
            \end{array}
        \right)
    \right),
\end{displaymath}
where $S^k$ is a univariate dataset that preserves the $k$-th attribute of each event in dataset $S$.

For $CT^*$, we encode the number of patterns, the sum of their usages, the distribution of their usages over different patterns, and the original patterns:
\begin{displaymath}
    \begin{aligned}
        L\left(CT^*\right)= & L_N\left(~|~P^*~|~\right)+L_N\left(usage\left(P^*\right)\right)            \\
                            & +log\left(\begin{array}{c}
                ~|~usage\left(P^*\right)~|~ \\
                ~|~P^*~|~
            \end{array}\right)+\sum_{p_i\in P^*}L\left(p_i\right),
    \end{aligned}
\end{displaymath}
where $P^*$ is the set of all patterns in $CT^*$.
Considering that $|~P^*~|$ and $usage(P^*)$ can be zero, we define $L_N(0)=0$.
For a non-singleton pattern $p_i$, we encode the number of events, the number of values, the number of gaps, the number of misses, and the first column (i.e., each value in the pattern) as
\begin{displaymath}
    \begin{aligned}
        L\left(p_i\right) = & L_N\left(~|~p_i~|~\right)+L_N\left(~||~p_i~||~\right)                            \\
                            & +L_N\left(gaps\left(p_i\right)+1\right)+L_N\left(misses\left(p_i\right)+1\right) \\
                            & +\sum_{v\in p_i}L\left(code_p\left(v~|~ST\right)\right),
    \end{aligned}
\end{displaymath}
where $L(code_p(v ~|~ ST))$ represents the encoded length of a value $v$ in pattern $p_i$.
Here, we directly use the pattern code of singleton value $v$ in $ST$.

\textbf{The encoded length of all code streams} is the sum of the description length of four types of codes:
\begin{displaymath}
    \begin{aligned}
        L\left(S~|~P\right)= & \sum_{p_i \in CT}usage\left(p_i\right)L\left(code_p\left(p_i\right)\right)     \\
                          & + \sum_{p_i \in CT}gaps\left(p_i\right)L\left(code_g\left(p_i\right)\right)    \\
                          & + \sum_{p_i \in CT}fills\left(p_i\right)L\left(code_f\left(p_i\right)\right)   \\
                          & + \sum_{p_i \in CT}misses\left(p_i\right)L\left(code_m\left(p_i\right)\right).
    \end{aligned}
\end{displaymath}

\end{document}